\documentclass[%
 reprint,
 amsmath,amssymb,
 aps,
]{revtex4-2}

\usepackage{graphicx}
\usepackage{dcolumn}
\usepackage{bm}
\usepackage[caption=false, justification=centerlast]{subfig}
\usepackage{lipsum}
\usepackage{graphicx}
\usepackage{comment}
\usepackage[mathlines]{lineno}
\usepackage{xcolor}
\usepackage{url}

\def\beq{\begin{equation}} 
\def\eeq{\end{equation}} 
\def\beqar{\begin{eqnarray}} 
\def\eeqar{\end{eqnarray}}

\def\avg#1{\langle #1 \rangle}

\def \nn{\nonumber}

\def \l({\left(}
\def \r){\right)}

\def \tu#1{\tau_{\rm #1}}
\def \ph#1{\phi_{\rm #1}}
\def \nc{n_{\rm comp}}

\def \delt#1{\Delta_{\rm #1}}

\def\avg#1{\langle #1 \rangle}
\def\lavg#1{\bigg\langle #1 \bigg\rangle}

\def\sig#1{\sigma_{\rm #1}}

\def\x#1{x_{\rm #1}}
\def\y#1{y_{\rm #1}}
\def\ph#1{\phi_{\rm #1}}

\setcitestyle{numbers,square}
\begin{document}
\preprint{AAPM/123-QED}

\title[Sample title]{Using Mutual Information to measure Time-lags from non-linear processes in Astronomy}  

\author{Nachiketa Chakraborty}
\affiliation{Data Assimilation Research Centre, Department of Meteorology, University of Reading, UK}
\author{Peter Jan van Leeuwen}
\affiliation{Department of Atmospheric Science, Colorado State University, USA, and Data Assimilation Research Centre, Department of Meteorology, University of Reading, UK
}%

\date{\today}

\begin{abstract}
Measuring time lags between time-series or lighcurves at different wavelengths from a variable or transient source in astronomy is an essential probe of physical mechanisms causing multiwavelength variability. Time-lags are typically quantified using discrete correlation functions (DCF) which are appropriate for linear relationships. However, in variable sources like X-ray binaries, active galactic nuclei (AGN) and other accreting systems, the radiative processes and the resulting multiwavelength lightcurves often have non-linear relationships. For such systems it is more appropriate to use non-linear information-theoretic measures of causation like mutual information, routinely used in other disciplines. We demonstrate with toy models the loopholes of using the standard DCF and show improvements when using  the mutual information correlation function (MICF). For non-linear correlations, the latter accurately and sharply identifies the lag components as opposed to the DCF which can be erroneous. Following that we apply the MICF to the multi-wavelength lightcurves of AGN NGC 4593. We find that X-ray fluxes are leading UVW2 fluxes by $\sim 0.2$ days, closer to model predictions from reprocessing by the accretion disk than the DCF estimate. The uncertainties with the current lightcurves are too large though to rule out -ve lags. Additionally, we find another delay component at $\sim -1 $ day i.e. UVW2 leading X-rays consistent with inward propagating fluctuations in the accretion disk scenario. This is not detected by the DCF. Keeping in mind the non-linear relation between X-ray and UVW2, this is worthy of further theoretical investigation. From both the toy models and real observations, it is clear that the mutual information based estimator is highly sensitive to complex non-linear correlations. With sufficiently high temporal resolution, we will precisely detect each of the lag features corresponding to these correlations.  
\end{abstract}

\keywords{correlation, time-lags, information theory, accretion, active galaxies}
\maketitle

\section{\label{sec:level1} Introduction}

Time domain astronomy has seen an upsurge in efforts and progress both in terms of methodology and results in providing insights into physical mechanisms as well as forecasting of transients. Correlations between multiple wavelengths and other messengers forms an essential tool in determining the geometrical configuration of variable sources and how transient mechanisms play out in them. \citep{1993PASP..105..247P}. They are used to extract time-lags between different wavelengths \citep{2009MNRAS.397.2004A}.  In accreting systems as active galactic nuclei (AGN), X-ray binaries (XRB), etc., these lags are attributed to reprocessing of variable high energy emission \citep{1991ApJ...371..541K}. Time lags are an important indicator of cause and effect within multiwavelength astronomy. They are also critical to high energy phenomena for constraining quantum gravity models leading Lorentz Invariance Violation (LIV) \citep{1998Natur.393..763A, 2013APh....43...50E}. When there is not one, but multiple mechanisms or models which are apriori equally plausible, model degeneracies might occur whereby it is critical to correctly identify and quantify time lags. 

Discrete correlation functions (DCF) \cite{1988ApJ...333..646E} have been the favoured tool to determine time lags from lightcurves. Backed by strong theoretical models, this approach has seen success in establishing lags \cite[for eg.,][]{2006MNRAS.367..801A, 2018MNRAS.480.2881M} as a function of physical parameters in the model. However, an assumption central to this approach is that the relationship between the variables being correlated is actually linear. When this assumption is valid and we have a single, strong mechanism that is apriori plausible as a causal link, the correlation studies ubiquitous in time domain astronomy will succeed in giving accurate results. 

There are several estimation challenges in this approach due to limitations in data such as sampling rate, statistics, as well as implementation challenges due to model dependence \cite{1988ApJ...333..646E, 1998PASP..110..660P} etc.  However, there is a more fundamental problem with cross correlation functions (CCF). If the relationship has a significant non-linear component, these correlation studies are not appropriate in general and the lags that they provide are likely to be erroneous \citep{pereda2005}. Accreting systems, which form a significant type of variable sources in astrophysics, are non-linear systems. Therefore they are susceptible to the aforementioned estimation problems. 

In this work we show with toy models how even for rather simple cases the lags computed are either inaccurate or not all lags are identified correctly. Furthermore, when there is not one, but multiple mechanisms or models which are apriori equally plausible, model degeneracies might occur whereby it is critical to correctly identify and quantify time lags. For example, for AGN there are multiple scenarios at play which leads to time delays between the X-ray emission and the longer wavelength emission at optical and ultraviolet (UV) parts of the spectrum. The most popular and successful scenario is where the X-rays produced closer to the central, compact source, are reprocessed in regions surrounding the central X-ray source and emitted at optical or UV wavelengths. The energy produced at these longer wavelengths will display a lag with respect to the X-rays as is clearly seen in several observations\citep{2008MNRAS.389.1479A, 2009MNRAS.397.2004A}. However, there is also the scenario where fluctuations in the accretion disk propagate from the outer region where there is long wavelength emission to the innermost regions, which produces X-rays. This non-linear propagation model, will produce UV/optical emission leading the X-rays, demonstrated in e.g. \citep{2001ApJ...550..655K, shemmer2003, 2008ApJ...677..880M} Therefore positive, negative  and indeed zero lags are plausible and often these scenarios are not seperated by a large time difference. Lags inferred from the observed time-series are very sensitive to the model differences and thereby prove vital in discriminating among them.

Information theory provides several measures which can be be applied to find time lags in systems with non-linear relationships. Measures as mutual information, transfer entropy, directed information, etc. have been used as causality measures 
in several other fields \citep[for eg.,][]{pereda2005, 2007PhR...441....1H, 2012Entrp..15..113A} including neuroscience, the geosciences, econometrics, etc. Most of these measures are based on evaluating conditional dependence. In other words, if the probability distribution functions of the time-series' in question are conditionally dependent in addition to presence of a time delay between them, then they are deemed to be linked causally. These measures take into account the full PDF of the variables as opposed to only the first few moments. Mutual information between two random variables, X and Y is the reduction in entropy of one, $H(X)$ given the other, expressed as $MI(X,Y) = H(X) - H(X|Y)$, and is a measure of dependence between these two time series. If X and Y are independent of each other, then $H(X|Y) = H(X)$ and the mutual information, $MI(X,Y) = 0$. Using Jensen's inequality on the definition of mutual information of two random variables, one can show that it is non-negative \cite{coverthomas}. 

In this paper we will study the behaviour of DCF in different toy problems and compare its performance in detecting time lags accurately with MI. We will show that MI is superior even when time series are linearly related because of its sharper discriminating power. Furthermore, we will apply both techniques on real lightcurves from  NGC 4593, where the MI will provide new insight into the physical mechanisms at play.

The paper is organised as follows. In section II the discrete correlation function and mutual information are introduced. In the section~\ref{sect:toymodels}, we investigate lags in simple sinusoidal toy models using both the DCF and our newly introduced Mutual Information Correlation Function (MICF). We explore 2 and 3 component models in sub-sections~\ref{subsect:twocompmodel} and ~\ref{subsect:threecompmodel} respectively. In the next section~\ref{sect:ngc4593mwl} we apply these tools to multiwavelength lightcurves of NGC 4593. Finally, in the section~\ref{sect:conc}, we discuss the conclusions and present our outlook towards future applications. 

\section{Causal measure : Mutual Information}
\label{sect:micf}

Typically, in time-domain astrophysics, relationships between two wavebands are established by linear correlation measures. This is then interpreted as a causal relationship in presence of plausible physical mechanisms, which the estimated correlation corroborates. And critical to establishing this causal relationship is the accurate quantification of temporal order of observed variables. The discrete correlation function is used to determine the temporal relationship quantified in terms of time-lags and was developed by Edelson and Krolik \cite{1988ApJ...333..646E} for application to unevenly sampled time-series in astronomy. This is computed from two time-series or lightcurves in terms of the pairwise as,
\beqar
\label{eqn:udcfedkr}
DCF( \tau) &=& \avg{x(t)y(t+\tau)} \nn \\ 
&=&\lavg{ \frac{(\x{i} - \bar{x})(\y{i} - \bar{y})}{\sqrt{\sig{x}^2-\delt{x}^2} \sqrt{\sig{y}^2-\delt{y}^2}} }
\eeqar
where $\sigma$ and $\Delta$ are the statistical standard deviation and the measurement uncertainty of the individual flux values (time-series variables : x and y) respectively. The angle brackets denote averaging over the pairs (i,j) of data points and would simply constitute an arithmetic mean of the pairwise correlations. We use the widely used implementation pyDCF \footnote{\protect\url{https://github.com/astronomerdamo/pydcf}}.

For non-linear systems, we define an analogous estimator in terms of mutual information between the two variables, as 
\beqar
\label{eqn:micf}
MICF(\tau) &=& MI( x(t), y(t+\tau) ) \nn \\
& = & \int p(x,y) \log \frac{p(x,y)}{p(x)p(y)}\;dxdy
\eeqar
where $p(x,y)$ is the joint probability density (pdf) of $x$ and $y$, and $p(x)$ and $p(y)$ are the marginal densities. The mutual information measures how far the joint pdf is from the product of the two marginals, so to what extent $x$ and $y$ are independent. Another way to express the mutual information is in terms of the Shannon entropy, H of the variables or random processes. It is given by, 
\beqar
\label{eqn:mi}
MI(x,y) &=& H(x) - H(x | y) 
\eeqar
Thus, the mutual information of $x$ and $y$ can be viewed as the reduction in Shannon entropy H of $x$ when $y$ becomes available. Since entropy is a measure of disorder, reduction of the entropy of $x$ by $y$ demonstrates that $y$ has information about $x$. If we consider the mutual information between lagged variable x and y, then a large mutual information at a certain time lag implies that the dependency constitutes a causal relationship. Thus, the mutual information correlation function (MICF) acts as an estimator of causality in presence of non-linear correlations. 

The MICF is calculated using the Kraskov et al. method \citep{2004PhRvE..69f6138K}. This is a non-parametric estimation of mutual information that avoids calculating the probability density functions but instead uses a k-nearest neighbours algorithm to evaluate the integrals directly. For each pair of lightcurves, we take a range of lag values from -ve to +ve and compute the mutual information between the lagged time-series ; the Kraskov method implicitly computes the average of nearest-neighbour mutual information contributions to give the MICF estimator.

\section{Toy model for multiplicative processes}
\label{sect:toymodels}
We want our toy models to capture important features of the underlying physics of the observed behavior. Observed time-series or lightcurves from AGNs and XRBs often show a linear relationship between the rms of the flux and its mean value for segments of the lightcurves \cite{2005MNRAS.359..345U}. This is one of the characteristics of a multiplicative process, which is by its very nature non-linear. Furthermore, numerous XRBs and AGNs show non-gaussian, in particular lognormal probability distribution functions for the observed fluxes, se e.g.  \citep{2005MNRAS.359..345U,2016A&A...591A..83S, galaxies8010007}, again suggesting a multiplicative process. While there are other ways of generating lognormal and other distribution functions compatible with the data, including linear processes \citep{2020ApJ...895...90S}, the multiplicative process is a natural way of describing aspects of the accretion disk physics and thus remains a compelling model. 

We will use such a multiplicative toy model as explained in \citep{1997MNRAS.292..679L} to illustrate the differences between predictions from the usual DCF and the mutual information function. As in \cite{2005MNRAS.359..345U}, a single time-series or lightcurve representing a multiplicative process described above can be expressed mathematically as:
\begin{equation}
\label{eqn:xt2comp}
    x(t) = \prod_{i} \left(1 + \sin\left(2 \pi i \nu \frac{t + \tu{i}}
    {N } + \ph{i}\right)\right)
\end{equation}

where $i\ (2\ \pi\ \nu )/ N$ is the angular frequency of the ith harmonic component, $\tu{i}$ and $\ph{i}$ are the time lag and phase respectively of the ith component. The time lags $\tu{i}$ represent direct differences in arrival times of photons, whereas $\phi$'s represent different components potentially arising from different mechanisms such as effects within the source itself \citep{Piran2005}. The $\phi$'s could also be effects due to particle physics or indeed other fundamental physics \citep{2005LRR.....8....5M, 2013APh....43...50E}. The details of the various competing mechanisms producing the time delays are beyond the scope of this paper. However, central to this work is that these mechanisms produce equivalent lags. While the intrinsic source term $\phi$ contains important physics we choose it equal to zero in our models, and absorb its influence in the time lags $\tau_i$.  

As mentioned above, we need to define the number of nearest neighbors in the MICF calculations. We tried a number of different values and found that the essential conclusion remains the same for these models. Figures 2 and 3, for the two and three component models, respectively, show the results of calculations for $k=5$. For a detailed study of dependence on k we refer the readers to \cite{Gao2014}. 

\subsection{Two component models}
\label{subsect:twocompmodel}
We start with the very simple case of two sinusoidal components or $\nc = 2$ for each of the two signals or lightcurves, x(t) and y(t) simulating 10 cycles of $N = 1000$ data points. We choose $\nu = 1/T = 1$ and only use one component $i=1$, leading to: 
\begin{eqnarray}
x(t) & = & (1 + \sin( 2\pi (t + \tu{x})/N )) + \eta \nonumber \\
y(t) & = &  (1 + \sin( 2\pi (t + \tu{y})/N )) + \eta
\end{eqnarray}
in which $\eta = 0.01*\mathcal{N}(0,1)$ is a small normally distributed noise component with zero mean and unit variance. We now choose the explicit time lags as $\tu{x} = 100$ for x(t) and $\tu{y} = 250$ for y(t).

\begin{figure*}
\subfloat[Time-Series (X,Y) and (Y vs X)]{\label{fig:sub1-2compdeltat150phase0}%
  \vspace{-2.0em}
  \includegraphics[width=.4\textwidth]{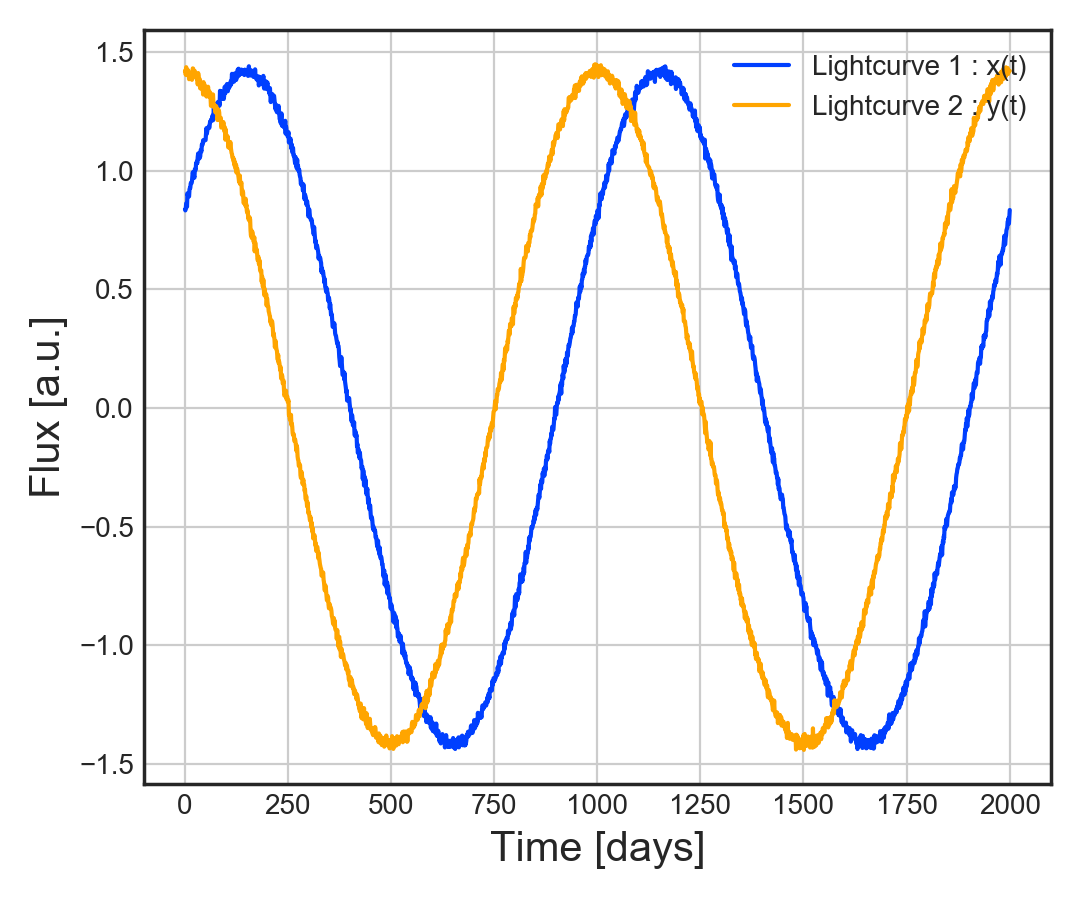}
  \includegraphics[width=.4\textwidth]{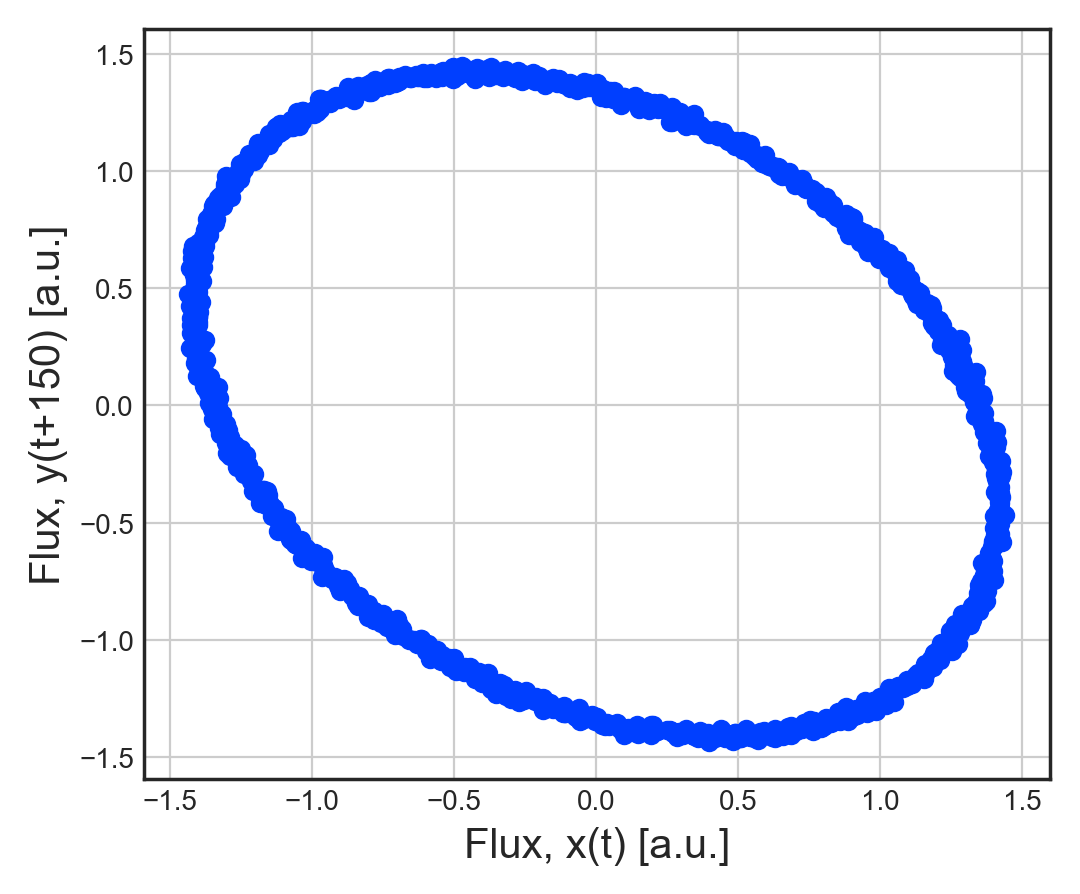}%
}%
\hfill
\subfloat[Lag Functions]{\label{fig:sub2-2compdeltat150phase0}
   \vspace{-2.0em}
   \includegraphics[width=.4\textwidth]{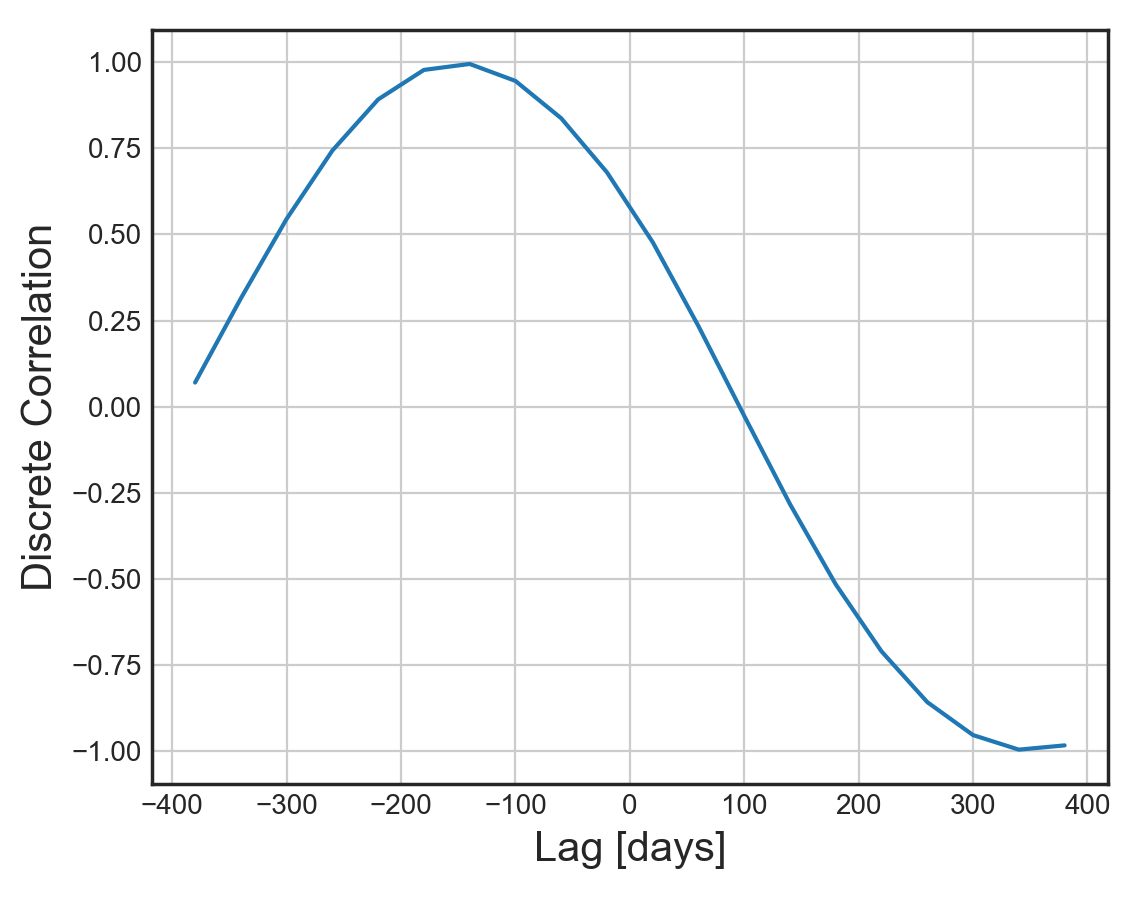}%
  \includegraphics[width=.4\textwidth]{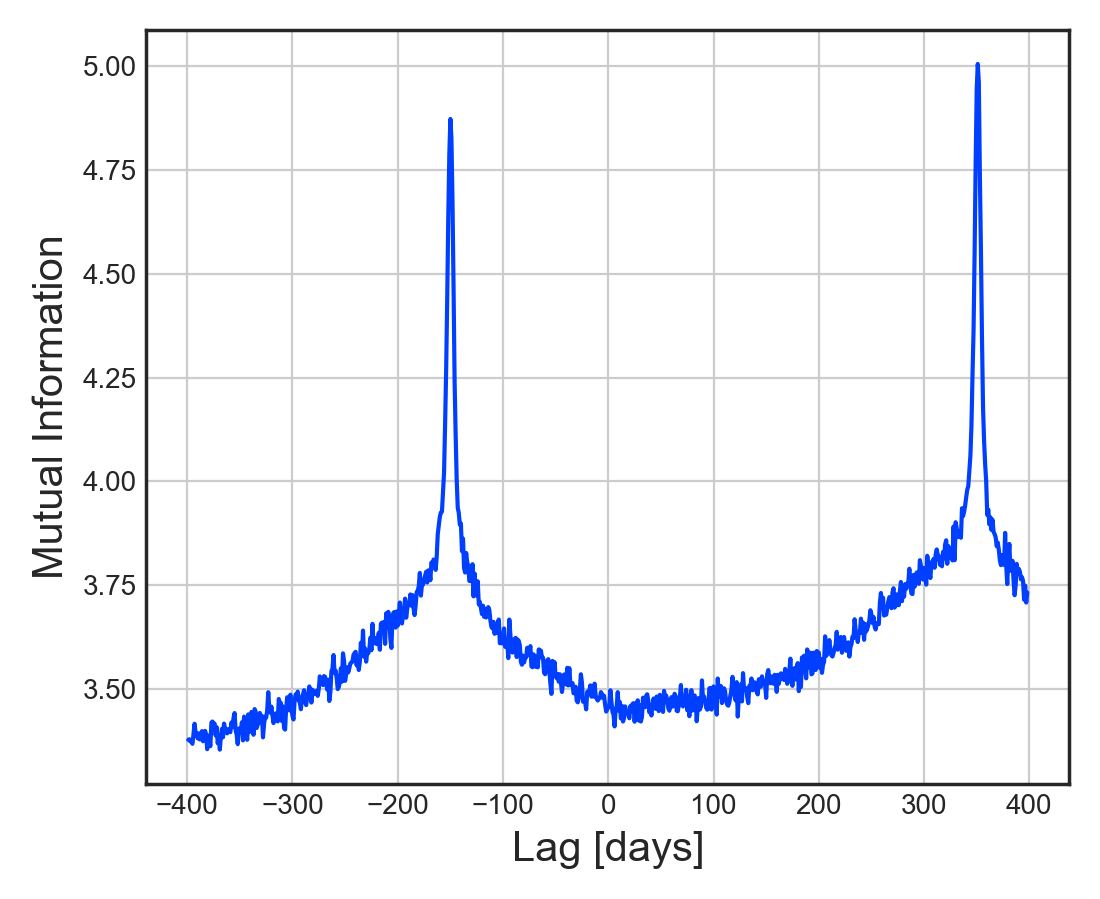}%
}
\caption[General Short Caption]{Top panel (a) shows the time-series of 2 component toy model with an explicit time difference of 150 lead for x(t) over y(t) and correlation plot for x(t) and y(t+150). Bottom panel (b) shows a comparison between the discrete correlation function (DCF) (Left:) and the mutual information correlation function (MICF) (Right:) for the lightcurves shown above. The lags are sharply identified by the MICF at the expected net lag (hence -ve) of 150 of y(t) and also for the lag corresponding to $\tu{x} + \tu{y}$ term at 350 . The former peak is not as sharply identified by the DCF and the latter is shown with a negative peak.\label{fig:2compdeltat150phase0}}
\end{figure*}

The timeseries and scatter plots of the two time series are shown in figure 1a, and the  DCF and mutual information correlation function, MICF in figure 1b. 
The correlation between x(t) and y(t) will have the term $ \sin( 2\pi (t + \tu{x})/N \times \sin( 2\pi (t + \tu{y})/N = 1/2 ( \cos(2\pi (\tu{x} - \tu{y})/N) - \cos(2\pi (\tu{x} + \tu{y})/N))$.  This gives the lags we see in figure~\ref{fig:sub2-2compdeltat150phase0} with MICF of $-150$ and $+350$ respectively. Thus, the MICF sharply identifies the correct lags with distinct peaks in the estimator. Compare this to the DCF, we find that a much broader peak appears at $-150$ and an equivalent trough appears for $+350$. The sinusoidal oscillatory functional dependence is not as sharp as the MICF. The profile of the peaks for clear identification is critical when there are several features as is evident in the case of the three component model in section~\ref{subsect:threecompmodel}. In fact, as we will see in the case of the different X-ray lightcurves in section~\ref{sect:4593}, this is true even for linear relationships between the correlating variables.

The MICF remains high between the two sharp peaks, suggesting a strong relation between $x$ and $y$ at all time lags. This is indeed the case, as can be seen in he scatter plot to the right of fig.~\ref{fig:sub1-2compdeltat150phase0} for a time lag of 150. In contrast, the DCF has a very small negative correlation at this lag as it tries to fit a line through these data points. This clearly demonstrates the advantage of the MICF over the linear measure such as DCF.

\subsection{Three component models}
\label{subsect:threecompmodel}
Now, we take a further step with a three component model. Once again, equation~\ref{eqn:xt2comp} in valid is general with $\nc = 3$. The frequencies are now at 0, $\pi / N$ and $2\pi / N$. For our test case we chose the identical time lags for each component as $\tu{x} = 200$ and $\tu{y} = 100$ respectively.  There are several interacting components in this case as there are multiple pairs in the three component model. We see this by expanding the product in x(t) and y(t) as,
\vspace{-0.6em}
\begin{align}
\label{eqn:xt3comp}
\hspace{-3.7em} x(t) = & [1 + \sin(\pi (t+\tu{x})/N) +\eta] [1 + \sin(2\pi (t+\tu{x})/N) + \eta]  \nn \\
= & 1 + \sin(\pi (t+\tu{x})/N) +  \sin(2\pi (t+\tu{x})/N) \nn \\
&+  1/2\cos(\pi (t+\tu{x})/N) - 1/2\cos(\pi (3(t+\tu{x}))/N) \nn \\
&+ \mathcal{O}(\eta) + \mathcal{O}(\eta^2) \nn \\
\hspace{-3.7em}y(t) =& [1 + \sin(\pi (t+\tu{y})/N) + \eta] [1 + \sin(2\pi (t+\tu{y})/N) + \eta] \nn \\
= &1 + \sin(\pi (t+\tu{y})/N) +  \sin(2\pi (t+\tu{y})/N) \nn \\
&+  1/2\cos(\pi (t+\tu{y})/N) - 1/2\cos(\pi (3(t+\tu{y}))/N)  \nn \\
&+ \mathcal{O}(\eta) + \mathcal{O}(\eta^2)
\end{align}

Once again, we add to each signal component a Gaussian noise component with zero mean and standard deviation 0.01, which is at a percent level compared to the signal.

\begin{figure*}
\subfloat[Time-Series (X,Y) and (Y vs X)]{\label{fig:sub1-3compdeltat150phasepiby2}%
  \includegraphics[width=.4\linewidth]{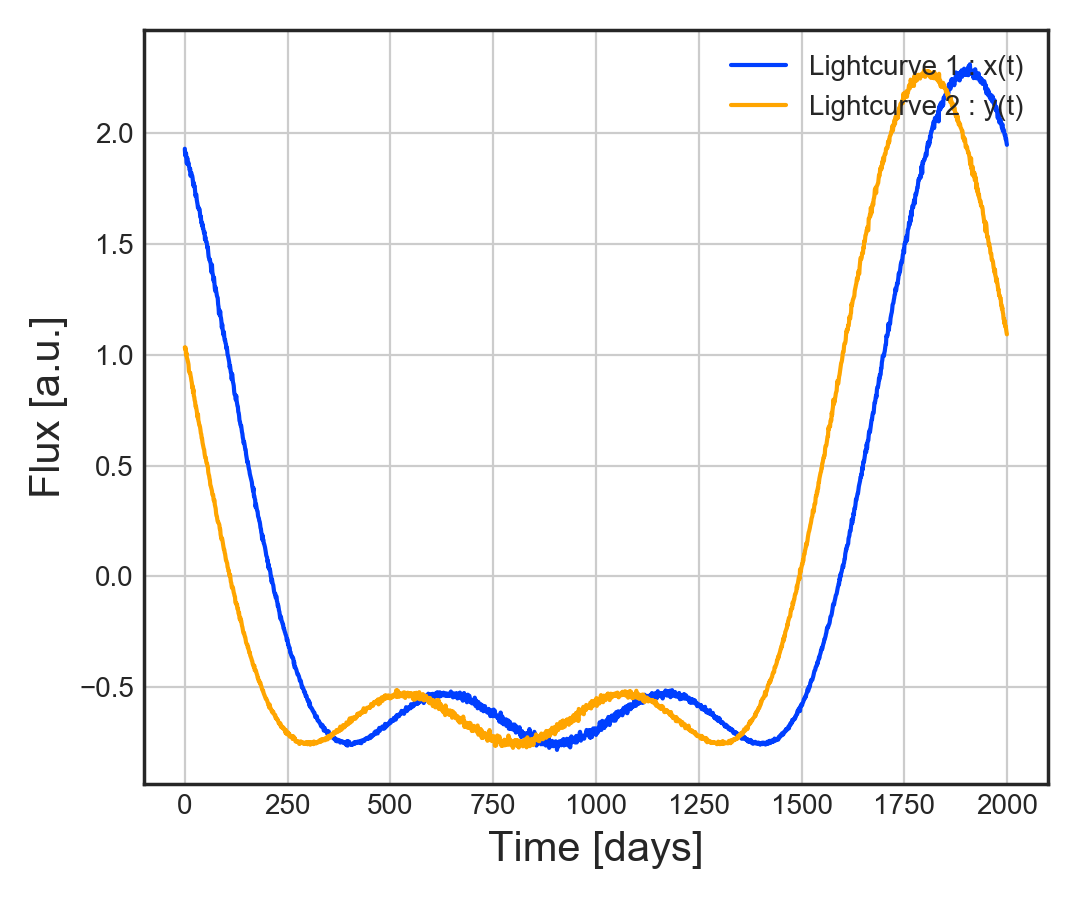}
  \includegraphics[width=.4\linewidth]{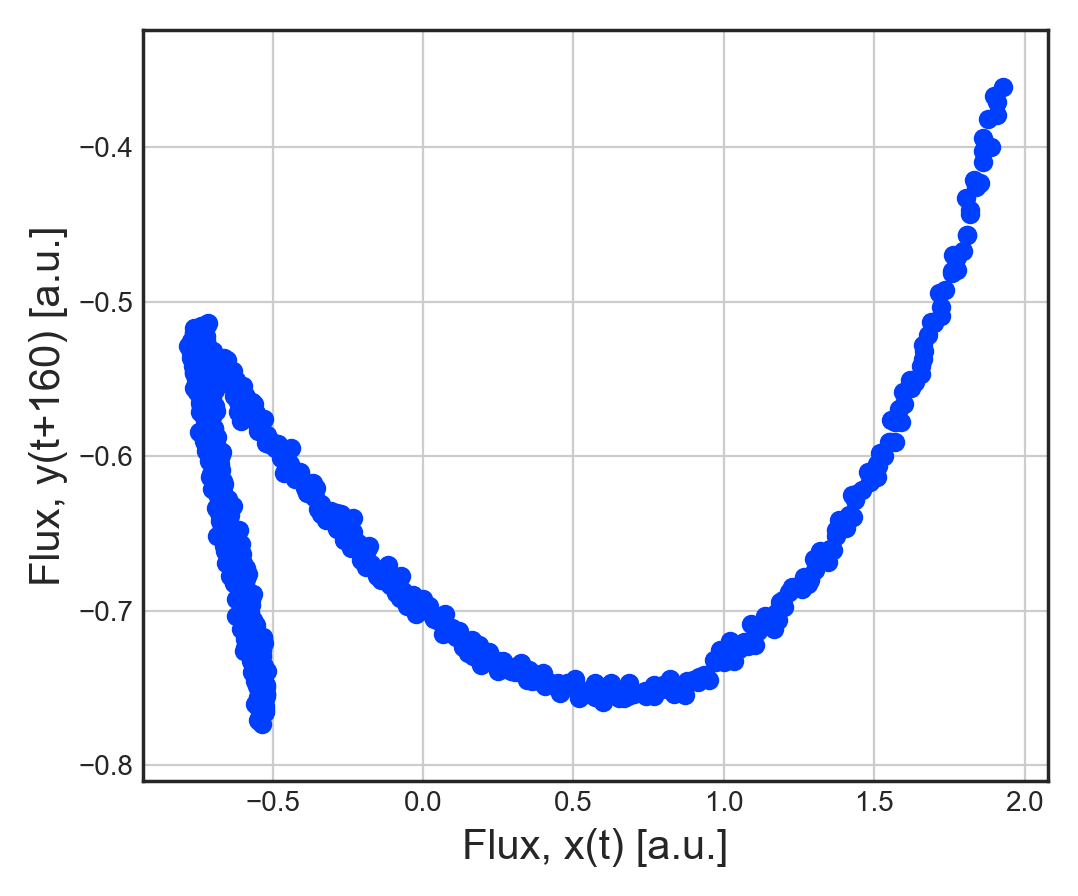}%
}%
\hfill
\subfloat[Cross Correlations]{\label{fig:sub2-3compdeltat150phasepiby2}%
    \includegraphics[width=0.4\linewidth]{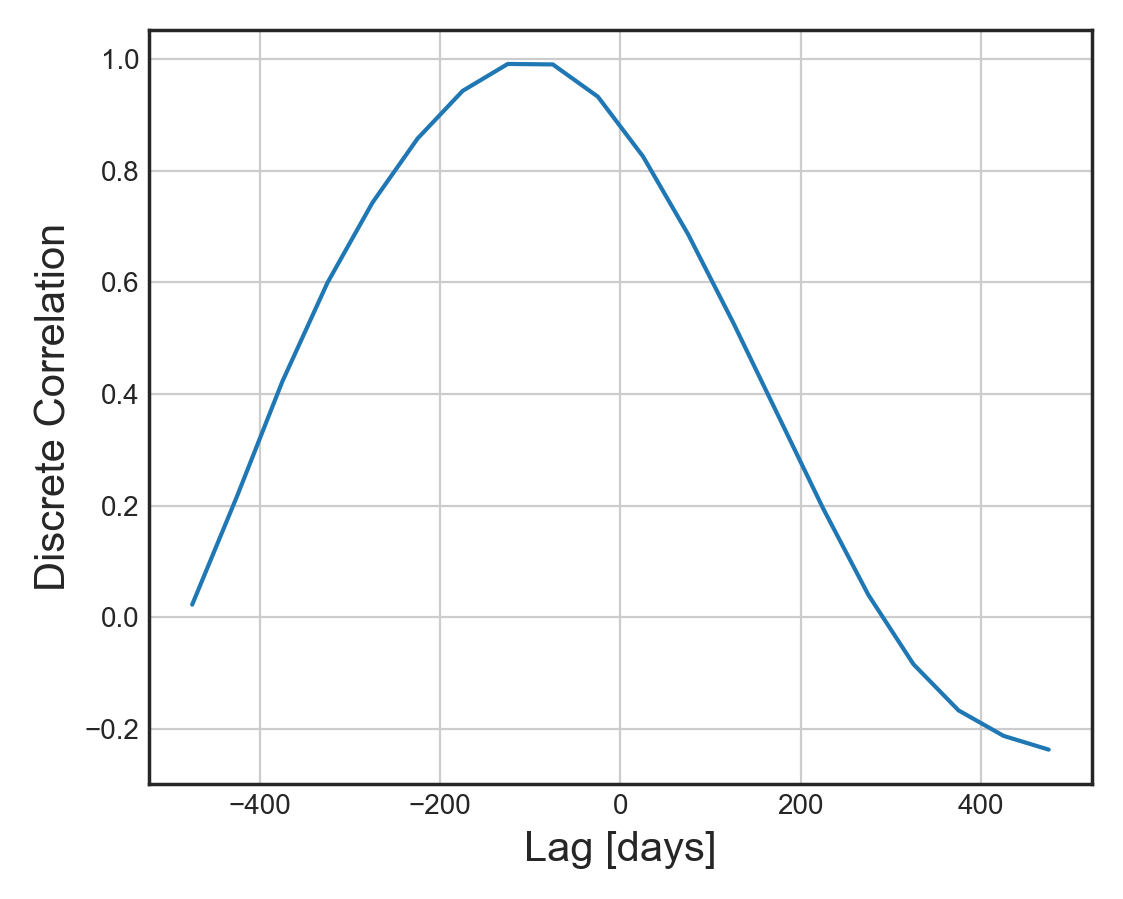}/
    \includegraphics[width=0.4\linewidth]{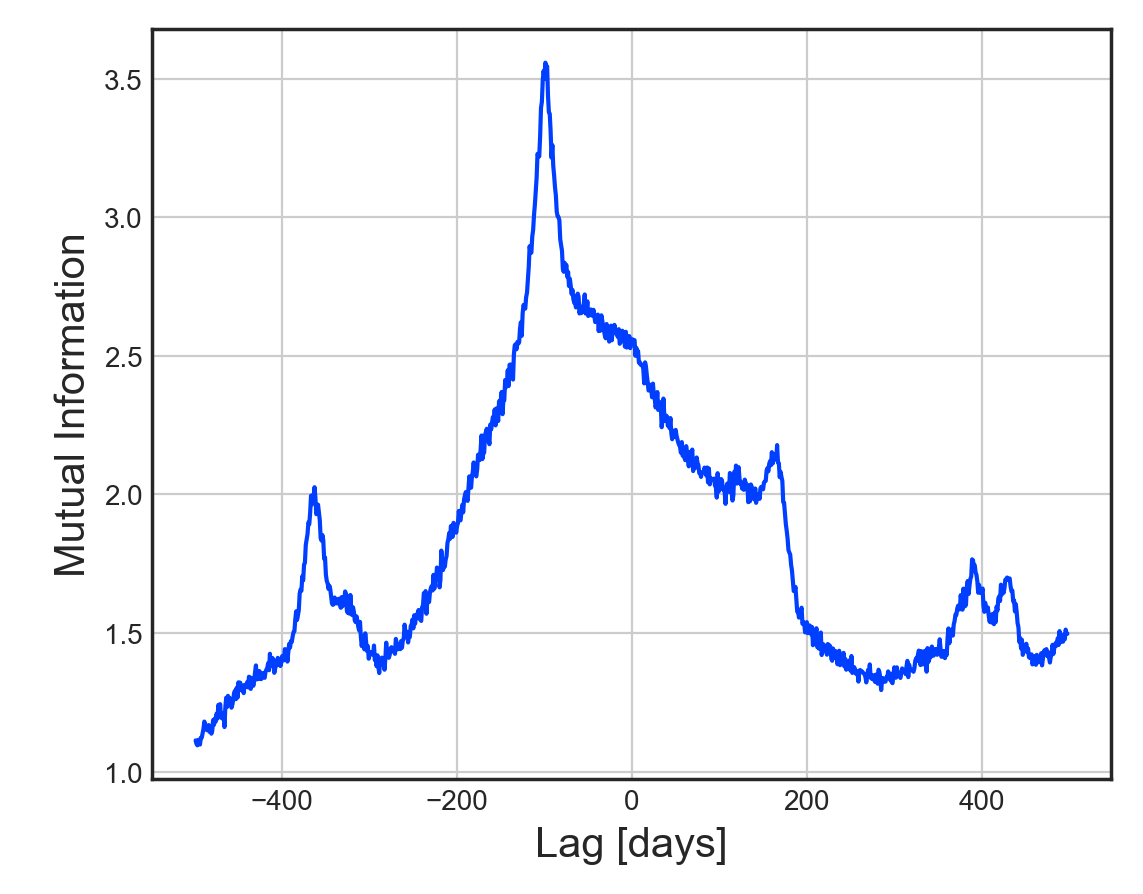}%
}
\caption[General Short Caption]{Top panel (a) shows a 3 component toy model with an explicit time difference of $100$ lead for x(t) over y(t) and a phase difference of 0.0 as in the 2 component case. $y(t+160)$ vs $x(t)$ is shown in blue to the right. Bottom panel (b) is same in fig~\ref{fig:sub2-2compdeltat150phase0}. Multiple strong delay components are sharply identified by the MICF with the highest peaks at $-100$ and $150$ as expected from calculating the different terms. The DCF does identify the dominant component accurately and misses out on the other ones.\label{fig:3compdeltat150phasepiby2}}
\end{figure*}

The DCF in figure~\ref{fig:sub2-3compdeltat150phasepiby2} shows a broad peak at the dominant lag at -100, but the peak is much sharper for the MICF. Equations (6) and (7) show that we should also expect peaks at other time lags between $\sim+150$ and $+\sim175$. The MICF identifies these as peaks, and also finds peaks at -350 and $\approx$400. As in the previous subsection~\ref{subsect:twocompmodel}, the scatter plot between x and y for lag features is quite revealing. We pick $\tau = 160$ and plot y(t+160) vs x(t) to the right of figure~\ref{fig:sub1-3compdeltat150phasepiby2}. This clearly shows that a strong relationship exists between y(t+160) and x(t), albeit a rather complex and highly non-linear one. This type of relation is not identified by the linear DCF. 
This figure shows that the MICF provides much more detailed and accurate information about the underlying time series than the DCF.

\section{Multiwavelength lags in NGC 4593}
\label{sect:ngc4593mwl}

Motivated by the studies on the toy models, we apply the MICF to actual observations. NGC 4593 is an active galactic nucleus or AGN. It is a Seyfert 1 galaxy and therefore a low luminosity AGN. As a Seyfert, it has a quasar type nuclei but a visible host galaxy. Therefore, there are emissions both from the host galaxy that are typically in UV/optical wavelengths and also from the accretion disc in the X-ray band. There is variability observed in each of these bands from several sources. One plausible source of variability is deemed to be fluctuations in the thermal emission from the accretion disc. An alternative scenario is the possibility of X-ray emission from the central corona or high energy UV photons from the inner edge of the accretion disc traveling to the outer disc region then re-radiating the emission that is observed. Therefore, time lags between different wavelengths measure distances between emission regions and are critical to our understanding of the emission mechanisms as well as the geometry of temperature profile of the AGN. Typically the lags between X-rays and UV / optical is used to map this in a procedure called "reverberation mapping" \cite{1982ApJ...255..419B}.

Now the estimated lags between X-rays and optical and UV are on timescales of days \citep{2008MNRAS.389.1479A, 2009MNRAS.397.2004A}, but, in some cases, a few months \citep{Zu_2011}. Typically observations support the lagging of optical emission and UV relative to X-rays, the uncertainties are such that these lags could be consistent with zero. These scenarios are compatible with the reprocessing of X-rays to generate longer wavelength optical and UV emission. However, there are a number observations \citep{2001ApJ...550..655K, shemmer2003, 2008ApJ...677..880M} showing X-rays lagging the longer wavelengths. Such cases could occur from inward propagation of fluctuations from the outer (long wavelength emission) regions to the innermost (short wavelength emission) region. In general, a combination of these two components can produce -ve as well as +ve lags in a X-ray - UV/optical correlation for such transient sources. 

\subsection{DCF vs MICF : Estimate of lags in NGC 4593}
\label{sect:4593}
\begin{figure*}
\subfloat[Time-Series (X,Y) and (Y vs X)]{\label{fig:1sub2-softhardxray}%
  \includegraphics[width=0.4\linewidth]{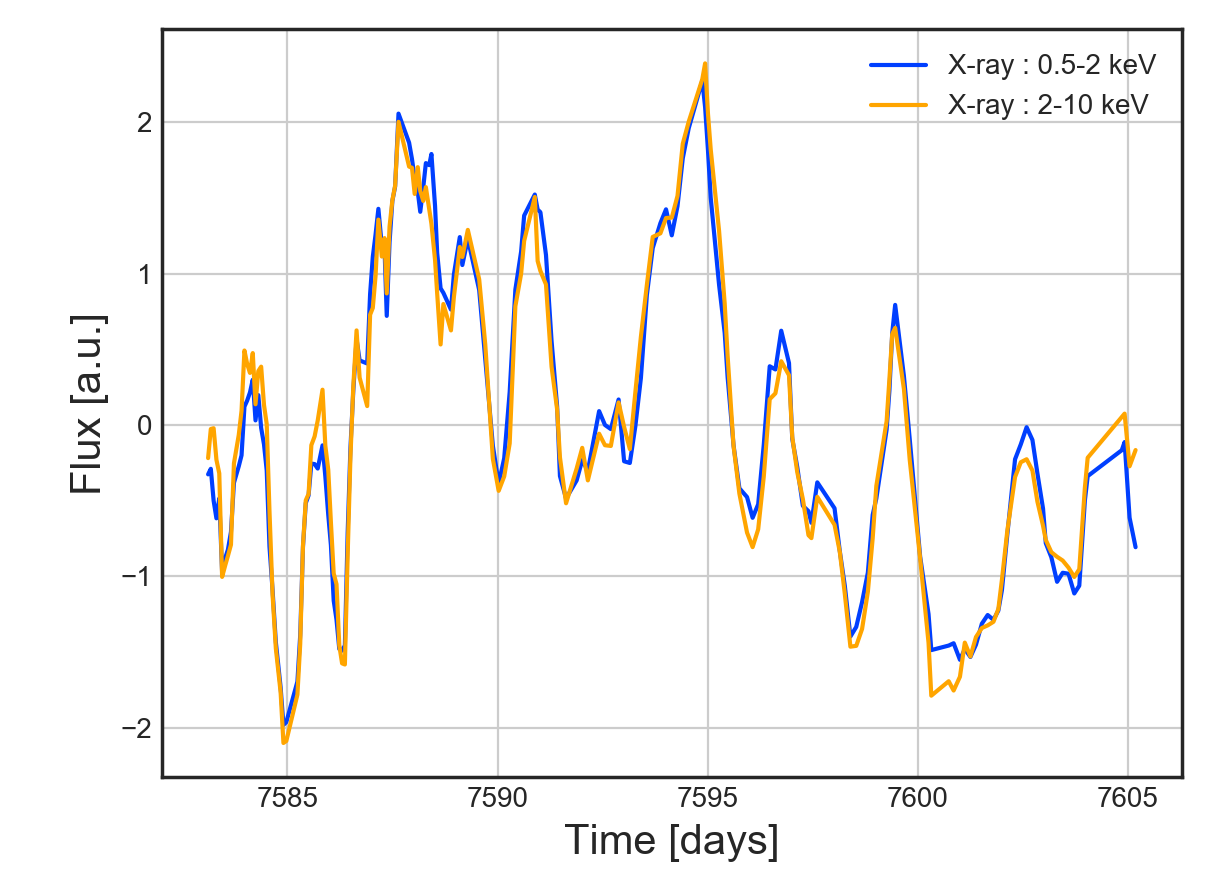}
  \includegraphics[width=0.4\linewidth]{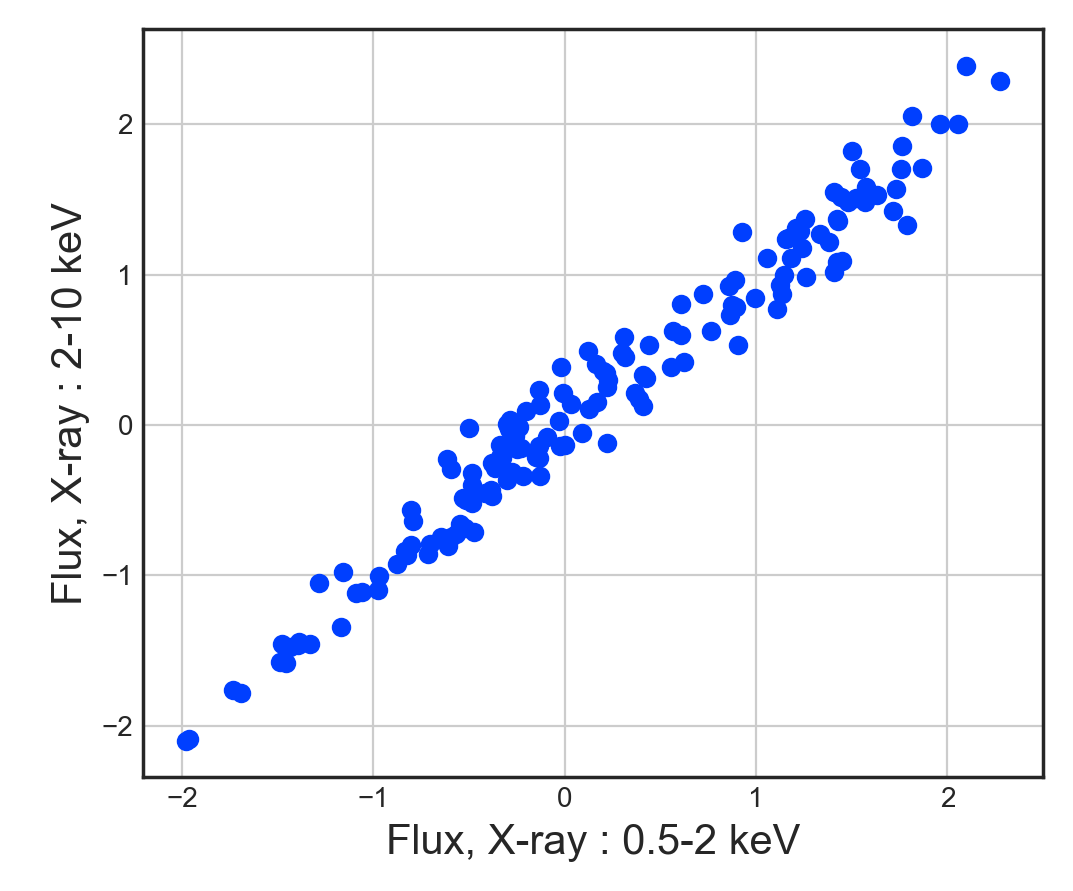}%
}%
\hfill
\subfloat[Cross Correlations]{\label{fig:2sub2-softhardxray}%
    \includegraphics[width=0.4\linewidth]{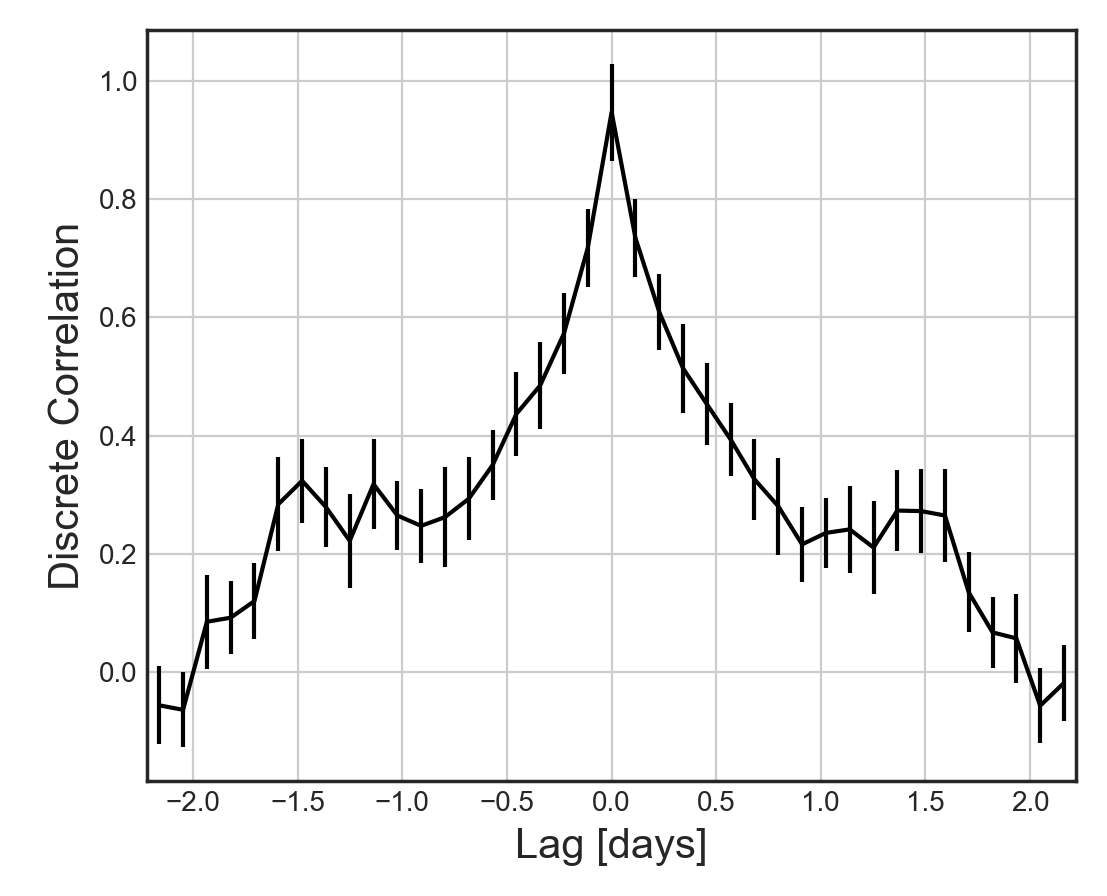}
    \includegraphics[width=0.4\linewidth]{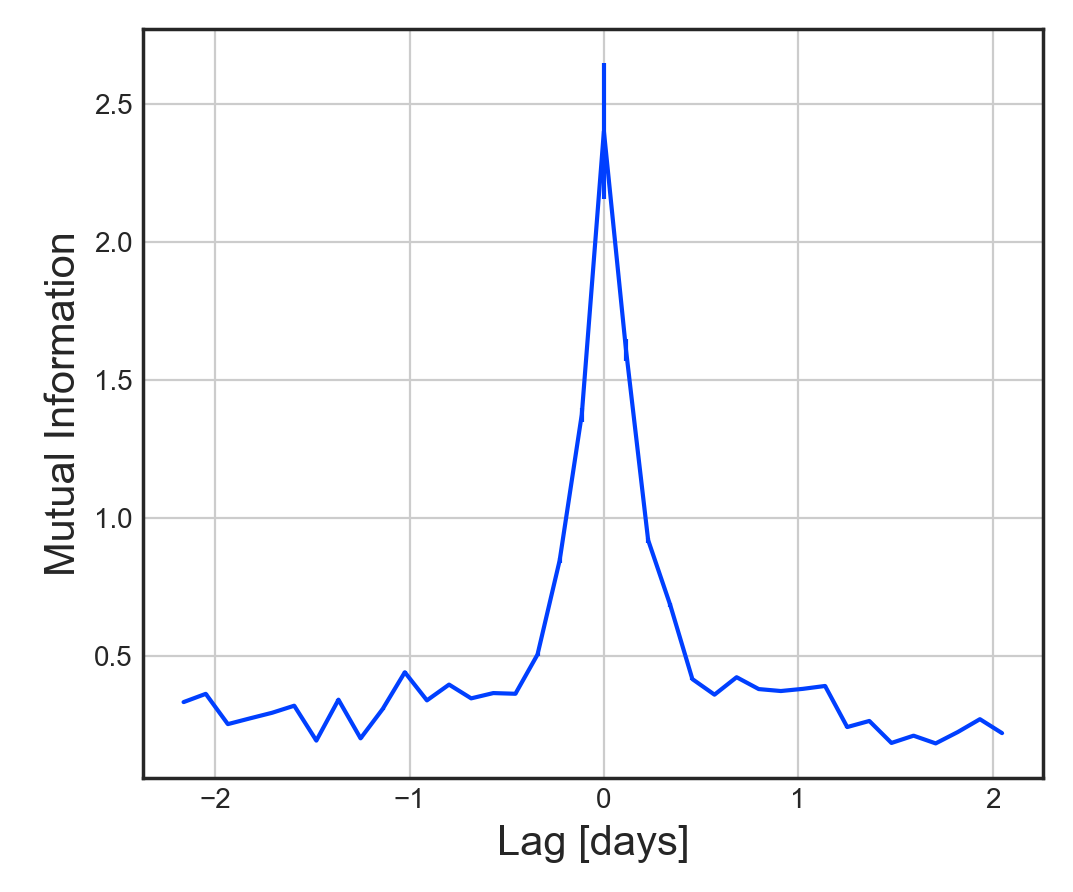}%
}
\caption[General Short Caption]{Top panel (a) shows soft (0.5-2 keV) and hard (2-10 keV) lightcurves (Left) and the flux-flux correlation plot (Right). Bottom panel (b) shows the In the bottom panel, the comparison between DCF and MICF is shown. There is agreement between the two on zero lag between the two X-ray bands. We see a much sharper identification with MICF noting the "shoulders" in DCF and the scales of the 2 subfigures.\label{fig:softhardxray}}
\end{figure*}

In the case of NGC 4593, \cite{2018MNRAS.480.2881M} performs a detailed analyses of these lags using the discrete correlation function aided by testing with simulated X-ray lightcurves. They estimate correlation of the X-ray lightcurves with other bands including the ultraviolet (UV) bands. We apply the MICF to these lightcurves for comparison with the DCF. We use $k = 5$ nearest neighbours in the Kraskov method as we did for the toy models, and the results are shown in figure 3. For these real observations as opposed to the toy models, we have noisy or stochastic lightcurves with uncertainties on individual flux values that maybe larger than in the toy models. Hence, we use smoothing to extract the key lag features over the stochastic variations. In order to do this, for MICF we use a rolling average with a window over 4 consecutive data points. We find that the detection of key features are robust to choice of window (as shown in Appendix~\ref{sect:appendixMI}). For the DCF, we simply use the slot weighting scheme that's built into the implementation of the Edelson and Krolik method, pyDCF \footnote{\protect\url{https://github.com/astronomerdamo/pydcf}}. This makes for a fair comparison between the two correlation estimators from less noisier signals. Identical range of lags are probed as shown in the figures~\ref{fig:2sub2-softhardxray} and ~\ref{fig:2sub2-xrayuvw2} for time increments equal to the median values of the bin (difference between the times of consecutive data points) of the unevenly sampled lightcurves. 

A key finding in \cite{2018MNRAS.480.2881M} is that the soft X-rays from 0.5-2 keV and the hard X-rays from 2-10 keV do not have any lag, contrary to previous findings. For this case of soft vs hard X-ray correlations, we use the same interval of lags $\approx [-2.2,2.2]$ with uniform time increments of 0.115 days which is the median interval for the X-ray lightcurves. We confirm the zero lag found by \cite{2018MNRAS.480.2881M} with our MICF, consistent with the strong linear correlation between the two bands with zero lag as shown in figure~\ref{fig:1sub2-softhardxray}. This linear correlation between the two X-ray bands is also shown in figure 4 in \cite{2018MNRAS.480.2881M}.

While there is an agreement  on the zero lag of the peak among the DCF and the MICF, the profile  is a lot sharper for the MICF. The DCF displays smaller secondary peaks or "shoulders" and then a further decrease in correlation for both positive and negative lags. In sharp contrast, the MICF has a steeper decline on either side of the peak with no "shoulders" or "wings". This is to be expected as the mutual information for linear relations is proportional to log(1-$\rho^{2}$). Therefore, we see that even for real datasets with a linear relation between two lightcurves, the MICF provides a sharper discrimination than the DCF. 

We next explore the relation between X-ray (0.5-10 keV) band and UVW2. In \cite{2018MNRAS.480.2881M}, the authors find a peak at $\approx 0.66$ days with the DCF. We can reproduce this result with the DCF. Our DCF estimate to the left in figure~\ref{fig:2sub2-softhardxray} looks for lag features in the range $\approx [-2.05,2.05]$ with a time increment of 0.107 days which is the median interval for the reference 0.5-10 keV lightcurve. As shown to the right of the figure~\ref{fig:2sub2-xrayuvw2}, we find one peak at $\approx 0.2$ days. We also find peaks at $\approx -1.1$ days. The positive lag is closer to the model prediction of $ \sim 0.1 $ days in \citep{2018MNRAS.480.2881M}. The negative lag supports the idea that also in this system lower energy information travels from the outer regions towards the inner accretion disk where it is transformed into higher energy emissions. The DCF is unable to pick this up clearly, and hence loses out on possible physical mechanisms at play. The MICF results seem plausible enough to be investigated further with astrophysical models.

\subsection{Uncertainty in lags in NGC 4593}

Because of measurement and sample noise and the short time series the dependency estimators are noisy. Uncertainty in the DCF is computed directly from the time-series or lightcurves as the standard error in the unbinned discrete correlation computed between pairs of points as in \cite{1988ApJ...333..646E}. From this we compute the uncertainty in the correlation, $d\rho$. We do not have an equivalent method for mutual information in general. One way to quantify their uncertainty is to generate extra artificial samples and perform significance tests. However, this can only be done reliably when the underlying statistical process is known ; in case of lightcurves of limited length as is the case here, estimates of such a process can be quite uncertain. A better way would be to divide the time series up in several segments and calculate an uncertainty estimate from mutual information calculations on each segment. Unfortunately, the time series are too short for a robust estimate. Instead, we will estimate uncertainty in the mutual information in the following way.

We base the uncertainty estimate on the assumption of Gaussian errors. The mutual information of two Gaussian random variables is given by $MI_{\rm g} = -1/2\ ln(1-\rho^2)$ where $\rho$ is the correlation between the two variables. We first calculate the mutual information using the Kraskov method as before. From that, we determine a corresponding artificial correlation value $\rho$ using the analytical expression above. We then determine the corresponding error in this correlation using the error determined for the DCF for this correlation value. Finally, we propagate this error in the artificial correlation back to the mutual information, using $dMI_{\rm g} = \rho/(1-\rho^2)\;d\rho$. This approach makes the uncertainty estimate consistent with the central value of the mutual information and the only approximation is that uncertainty calculation assumes Gaussian errors.  

In this manner, we estimate the uncertainty in the features in both the DCF and the MICF. For the soft vs hard X-ray correlation, the peak at zero lag in figure~\ref{fig:2sub2-softhardxray} has a discrete correlation of $0.95\pm0.08$. This value falls to half its value at lag values $\approx \pm 0.6$ days (i.e. full width half maximum $\approx 1.2$ days). A more precise estimate of the uncertainty in position (or lag value) of the central peak obtained by fitting the DCF in the figure with a Gaussian, yields an error of $0.53$ days. The corresponding peak in the MICF at zero lag has a mutual information value, $MI = 2.40\pm0.25$. The uncertainty at the central peak is much larger that for the other values, which are barely visible on the scale in figure~\ref{fig:2sub2-softhardxray} owing to this large value of correlation $\rho$. The mutual information falls to half its value well within $\approx \pm 0.2$ days. Fitting the MICF in the figure with a Gaussian gives an error of $0.23$ days. Thus, the MICF has a sharper peak as explained before. 

\begin{figure*}
\subfloat[Time-Series (X,Y) and (Y vs X)]{\label{fig:1sub2-xrayuvw2}%
  \includegraphics[width=.4\linewidth]{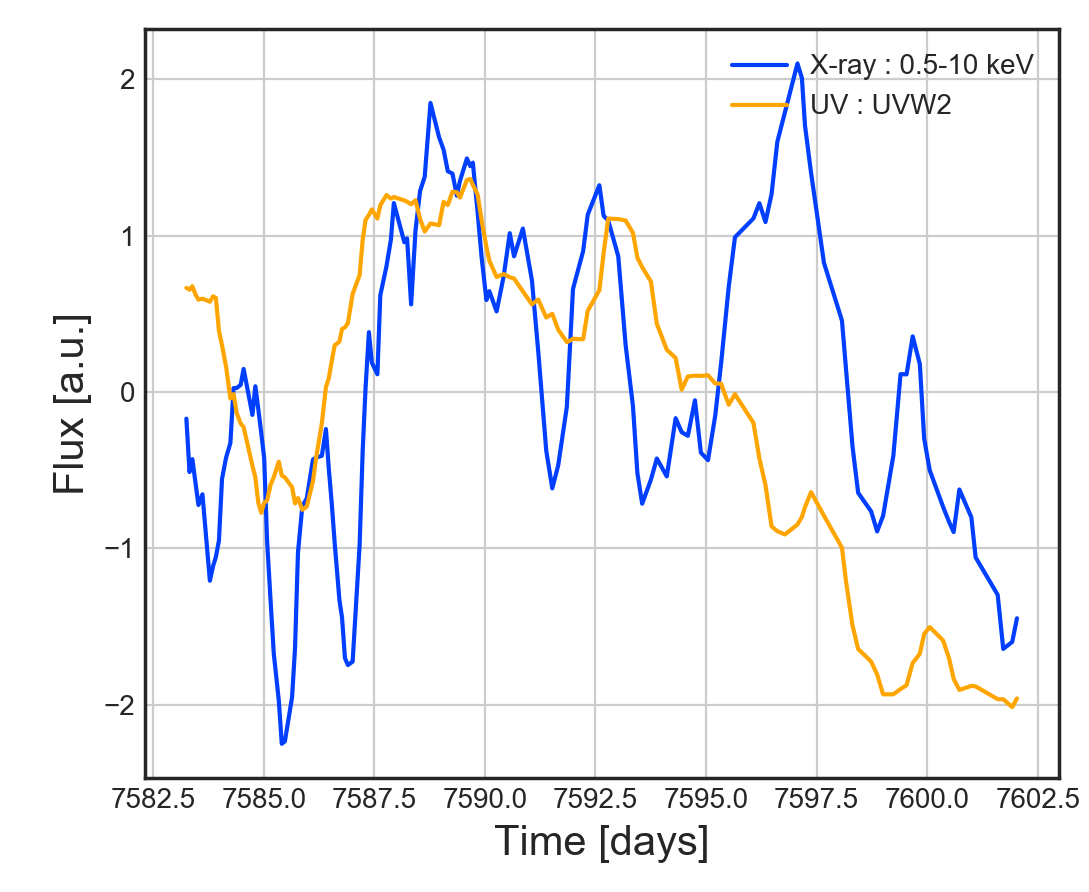}
  \includegraphics[width=.4\linewidth]{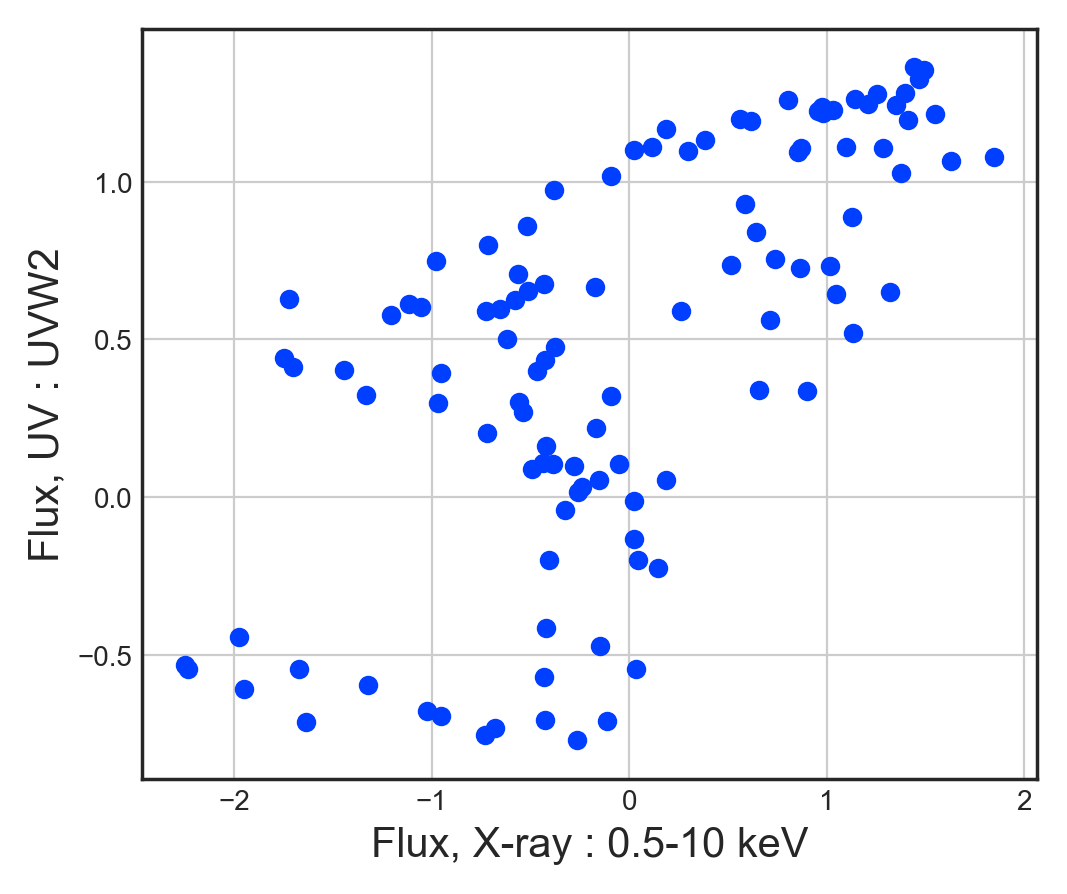}%
}%
\hfill
\subfloat[Cross Correlations]{\label{fig:2sub2-xrayuvw2}%
    \includegraphics[width=0.4\linewidth]{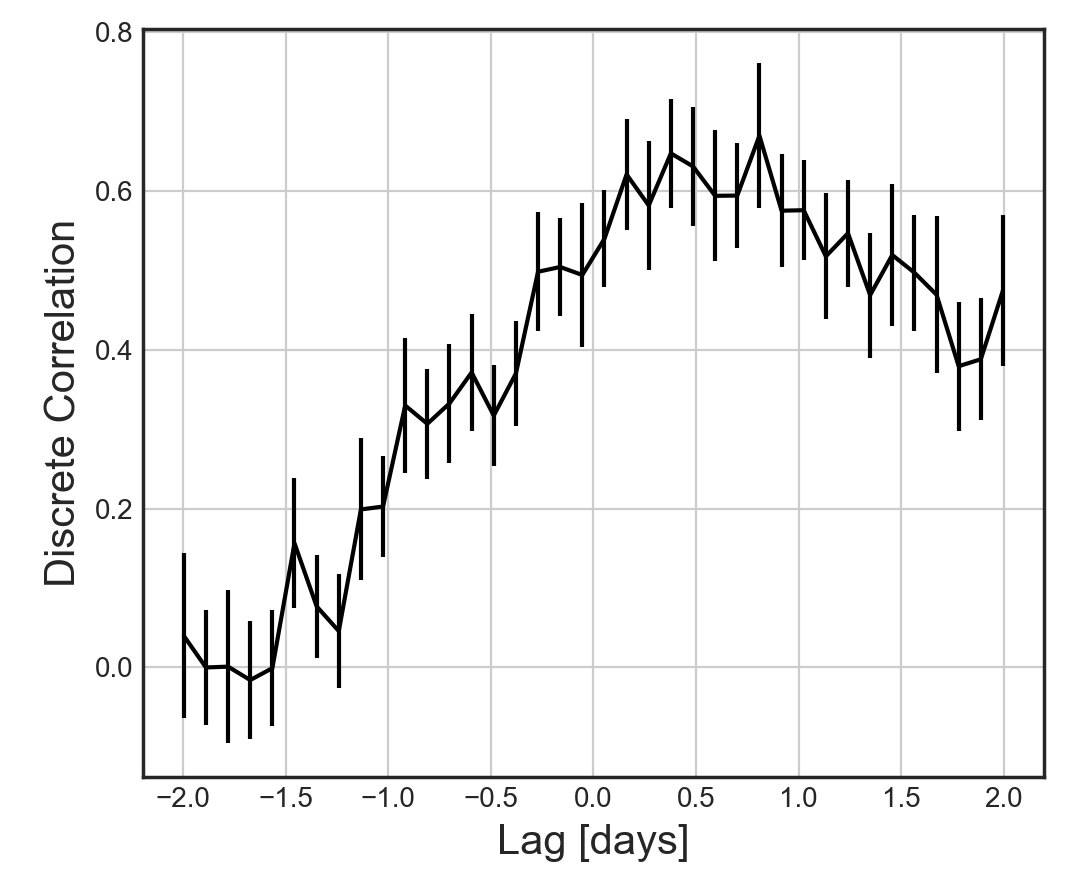}
    \includegraphics[width=0.4\linewidth]{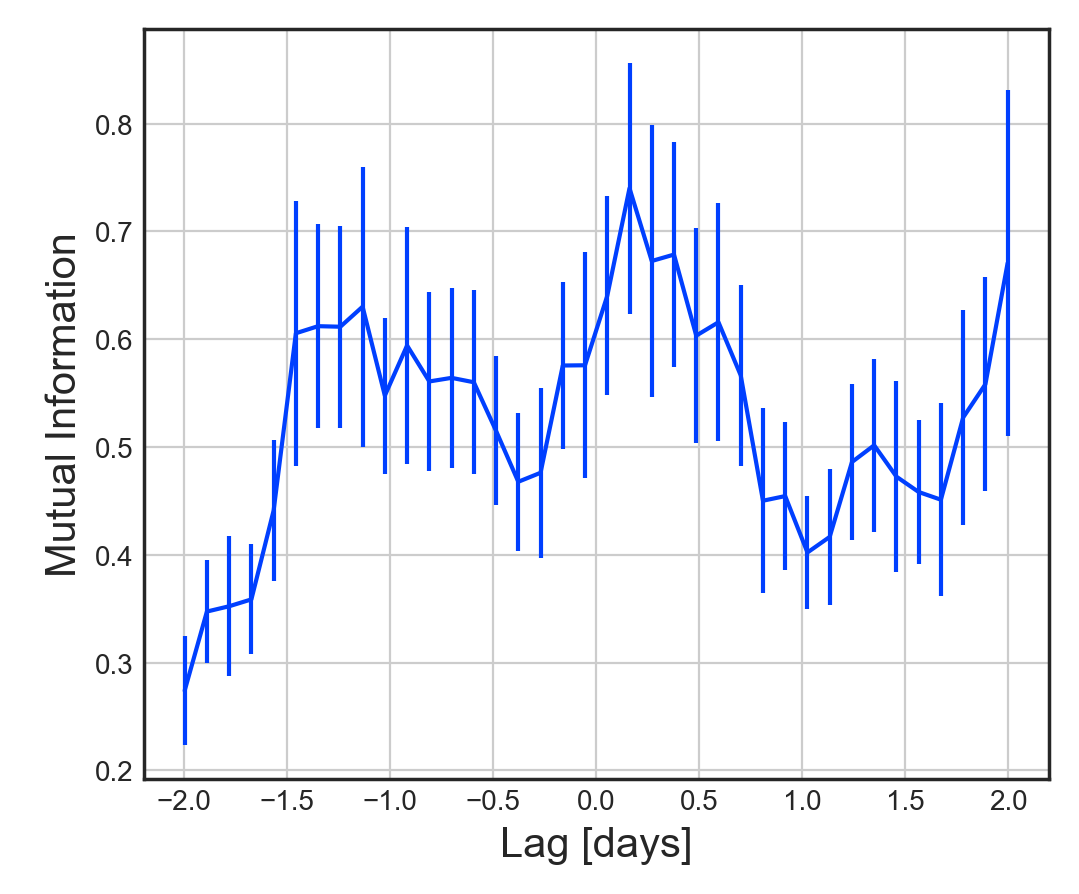}%
}
\caption[General Short Caption]{Top panel (a) shows X-ray (0.5-10 keV) lightcurves in blue and UVW2 in orange (Left) and the flux-flux correlation plot (Right). Bottom panel (b) shows the comparison between DCF and MICF. The DCF shows a peak at a lag of $\gtrsim 1/2$ day roughly consistent with reported value. It doesn't show any other prominent peak. The MICF also shows its highest peak at $\approx 0.2$, but shows another competitive peak $\approx-1.1$. The uncertainties on mutual information values are computed by propagating the uncertainty on the corresponding $\rho$ from DCF analytically.\label{fig:xrayuvw2}}
\end{figure*}

For the X-ray-UV correlation, \cite{2018MNRAS.480.2881M} finds a peak at 99$\%$ confidence at $0.66\pm0.15$ days. It evaluates significance levels by simulating the reference X-ray lightcurves. As we explain earlier in the section, we take a different approach. While we can reproduce their results, we use a time increment of $0.107$ days or the median value with slot (uniform) weighting to produce figure~\ref{fig:2sub2-xrayuvw2}. This gives us a central value of $0.38$ days. The peak value of the discrete correlation is $0.64\pm0.05$. The uncertainty estimate is more challenging here with a rather broad and asymmetric profile. A conservative estimate from the -ve lag value at which the DCF falls to half its peak value yields an uncertainty of $\approx 0.87$ days. Moreover, as the MICF reveals, there are multiple features that need to be disentangled. Using the sensitivity of the MICF, we are able to isolate the lag centered around $\tau \approx 0.16$ days with a 4 point window to the right of figure~\ref{fig:2sub2-xrayuvw2} with mutual information, $MI = 0.74\pm0.12$. The mutual information diminishes as we move away though not down to half its peak value as we start to see a rise especially as we draw closer to the second feature at $\approx -1.1$ days. For the uncertainty on the lag at $0.16$ days, we fit the segment of the MICF between the two local minima either side of this peak with a Gaussian. This gives us a uncertainty estimate of $\approx 0.73$ days. Fitting this feature for different windows (no window, 2 and 4 in section~\ref{sect:appendixMI}) gives us an error estimate ranging from $\approx 0.6 - 0.8$.   

\section{Discussions and Conclusions}
\label{sect:conc}

Mutual information is a non-linear measure to quantify the statistical dependency between two or more  variables. Therefore it is more appropriate than the cross correlation to quantify dependency between variables that deviate from a linear relation. This is clearly demonstrated by the toy models, which show that when the relation between lagged variables x(t) and y(t+$\tau$) is non-linear, the mutual information based function finds the corresponding lags while the standard discrete correlation does not.

Accreting astrophysical sources like AGNs and XRBs tend to show characteristics that can be naturally explained by multiplicative processes driving the variability of their emitted radiation \citep{1997MNRAS.292..679L, 2005MNRAS.359..345U}. This leads to different wavebands being non-linearly related. We demonstrate, with the help of an analytical toy model mimicking this multiplicative behaviour, the difference between lags computed with the standard discrete correlation function and with the mutual information correlation function or MICF. The latter correctly detects the lag features that are present in the toy models, some of which were missed by the DCF. The MICF also identifies the lags more sharply than the DCF, a feature also found when using real observations. 

Having shown this proof-of-principle with toy models and tested the simpler relationship between different X-ray bands, we apply MICF to the real observations. First we apply it to test the correlation between soft (0.5-2 keV) and hard (2-10 keV) X-rays. We confirm that the principal lag feature is centered symmetrically around zero consistent with the finding in \cite{2018MNRAS.480.2881M} with $\tau = 0.0\pm0.23$ days from the MICF. The uncertainty quantification shows a sharper identification of this strong feature. Then we move to the relation between X-rays and UV emitted by NGC 4593, which is the most interesting case physically. The standard reflection scenario predicts X-rays being reprocessed into and therefore leading the longer wavelengths such as the UV. For NGC 4593, the DCF shows a maximum at a positive time lag at approximately half a day. However, the correlation between X-rays and UV-W2 band is found to be non-linear. Hence, we use the more appropriate non-linear measure, the mutual information. We find that the MICF peaks at a lag, $\tau\approx0.16\pm0.73$ days. This central value of this peak is in close agreement with the predicted lag of $\sim 0.1$ day from the reprocessing model between UV and X-rays. However, the uncertainty is large enough that the lag could be consistent with zero or even have a negative value. Observations of a number of AGNs also show that UV can lead the X-rays in case of inward propagating fluctuations within the accretion disk. This scenario is not ruled out by our estimates. 

Indeed, we detect an additional lead of 1.1 days in the MICF, which could arise from inward propagating fluctuations. The DCF does not pick up this lead, which could be related to the strong nonlinearity of this propagation process.  Indeed, the X-ray flux vs UVW2 flux plot in figure~\ref{fig:1sub2-xrayuvw2} shows a strongly nonlinear relation between the two. This explains why the MICF would show a better agreement with the positive lag of UV with respect to X-rays predicted by the reprocessing model, and the negative lag at 1.1 days. However, the uncertainty estimates of current studies are large enough that absence of such leads cannot be ruled out, see e.g. \citep{2018MNRAS.480.2881M}, and further theoretical investigation is warranted.

This study shows the importance of non-linear measures of statistical dependency in establishing causal mechanisms from lightcurves. This is of utmost importance in astrophysics, where typically linear correlation measures are used to corroborate the models capturing the underlying causal mechanisms. From this study, it is evident that for even relatively simple non-linear relationships, linear cross-correlations can give inaccurate time-lags and correlations in general. This includes incorrect lag values as well as missing certain lags entirely. 

Therefore, we propose to use nonlinear lag-detection measures for observations where it is clear that there are strong non-linearities between the multi-wavelength fluxes. Furthermore, even when there is a linear relationship between the correlating variables, the inherent properties of mutual information enable a sharper identification of time-lags between the variables. In general, when there are multiple lag features the profile of the peaks is crucial in distinguishing them. This would be critical for accreting systems, where multiple processes can be at play producing different time delays \cite{2001ApJ...550..655K, shemmer2003, 2008MNRAS.389.1479A, 2009MNRAS.397.2004A, 2008ApJ...677..880M, Zu_2011}. If these features operate at similar timescales, a sharper resolution will allow to separate them. 

 Application of nonlinear measures is not limited to models of accreting systems but can be used in AGNs and gamma-ray bursts to constrain violation of Lorentz Invariance \citep{1998Natur.393..763A, 2005LRR.....8....5M}. Of particular interest would be cases of significant flares at gamma-ray energies that have enough photons \citep{2007ApJ...669..862A, 2013PhRvD..87l2001V, 2019ApJ...870...93A}. This gives us the opportunity to devise a sensitive detection of lags at timescales down to a few minutes would put strong constraints on the underlying quantum gravity models. 
 
We have shown and argued that the mutual information function is a good choice for the systems discussed above, although there maybe others that will be explored in future. The MICF is easy to compute and applicable to a wide range of conditions including continuous and discrete variables given the work done in \citet{2004PhRvE..69f6138K} and modifications such as \citet{2020arXiv201002247V}. Mutual information estimates do demand more and better quality data to provide the sharpest identification of complex features it is sensitive to. With longer lightcurves and improving quality of data in terms of cadence and signal to noise ratio, we expect better results for time-lags leading to strong constraints on the physics of these systems. 
\begin{acknowledgments}
Both authors are supported by the European Research Council under the Horizon 2020 program, via the CUNDA project number 694509. We acknowledge and thank Ian McHardy for providing the lightcurve data used in this study. 
\end{acknowledgments}

\appendix
\section{Effect of Smoothing on the lags from Mutual Information}
\label{sect:appendixMI}
\begin{figure*}
\subfloat{%
    \includegraphics[width=0.33\linewidth]{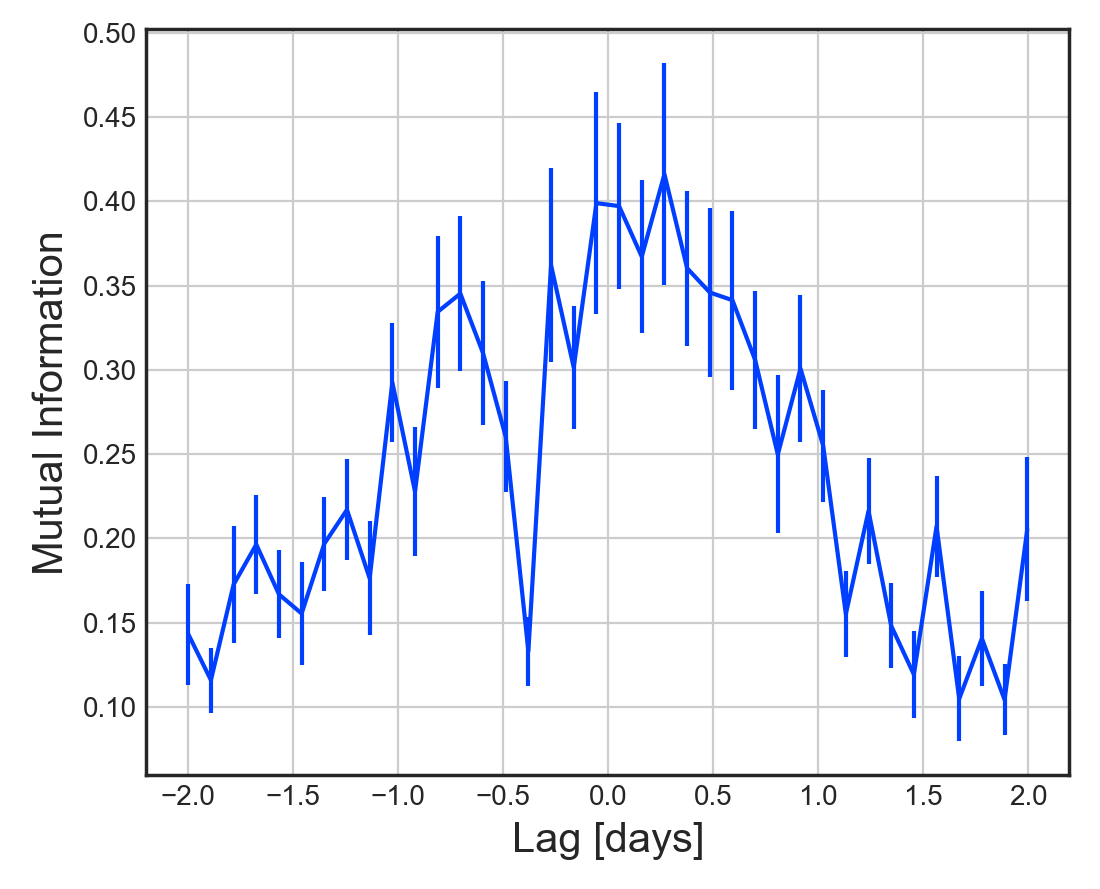}
    \includegraphics[width=0.33\linewidth]{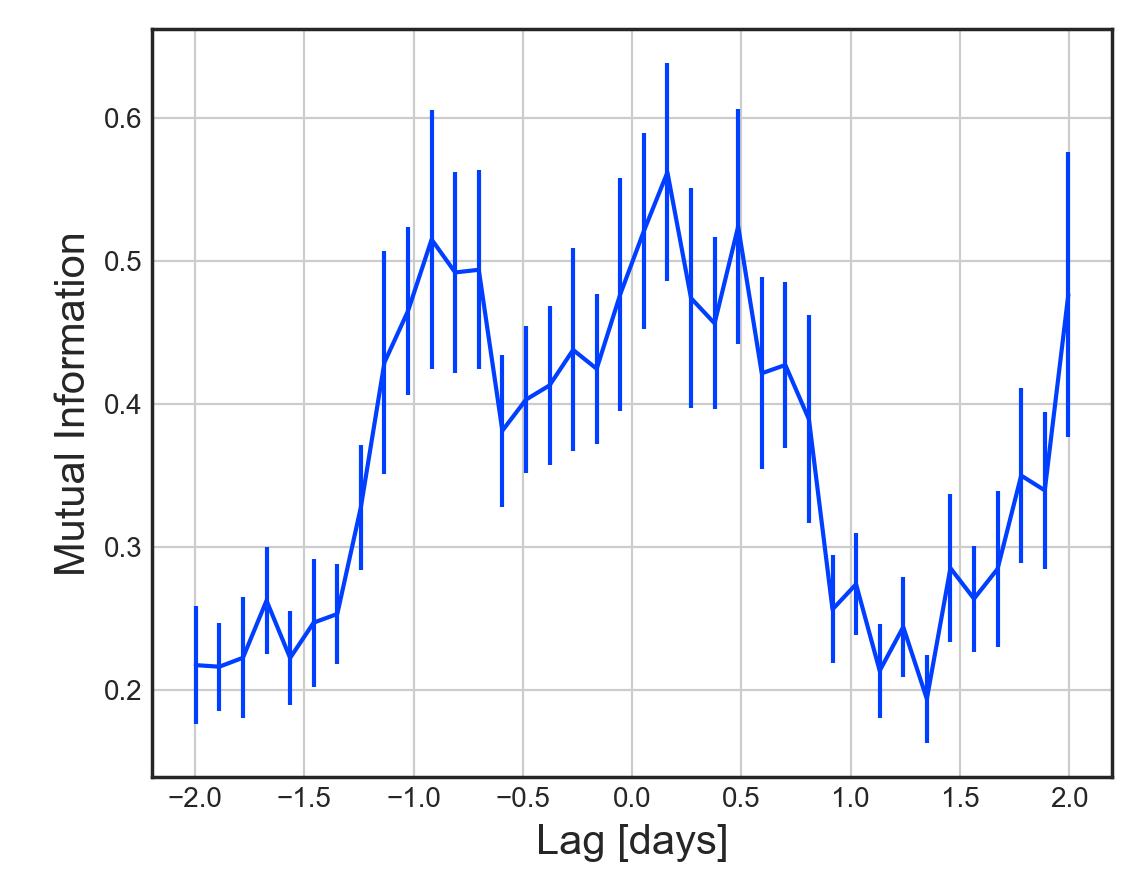}
    \includegraphics[width=0.33\linewidth]{figuresnewdata/seabornbright_micf_0p5to10_sigrhofromDCF_keq5_win4_nlag20_length148128_m2p05to2p05dt0p107slot.png}
}
\caption[General Short Caption]{The effect of the rolling window on the MICF estimate is shown here. The MICF with no window (Left:), one averaging over 2 consecutive points (Center:) and 4 consecutive points (Right:) all show peaks close to the predicted lag feature at 0.1 days.}
\label{fig:appendix-micfxrayuvw2}
\end{figure*}
On account of the noisiness of the observed lightcurves, we use smoothed lightcurves. For the MICF, we do this by using a rolling window. This allows to retain most of the data points in the lightcurve, even though this procedure does violate the assumption of independent and identically distributed (iid) samples within each window. However, it is clear from the figure~\ref{fig:appendix-micfxrayuvw2} that for each of these smoothed lightcurves, the estimated MICF shows a peak near the model prediction for the time lag of 0.1 days in \cite{2018MNRAS.480.2881M}. There is additional feature between $-0.5$ and $-1.5$ days where the X-rays are seen to lag the UV. In this case the window does alter the central value of this peak. This deserves further investigation. 

\section{Effect of Smoothing on the lags from Discrete Correlation}
\label{sect:appendixDCF}
\begin{figure*}
\subfloat{%
    \includegraphics[width=0.33\linewidth]{figuresnewdata/seabornbright_dcf_0p5to10_uvw2i_sigrhofromDCF_keq5_nlag20_length148128_m2p05to2p05dt0p107slot.png}
    \includegraphics[width=0.33\linewidth]{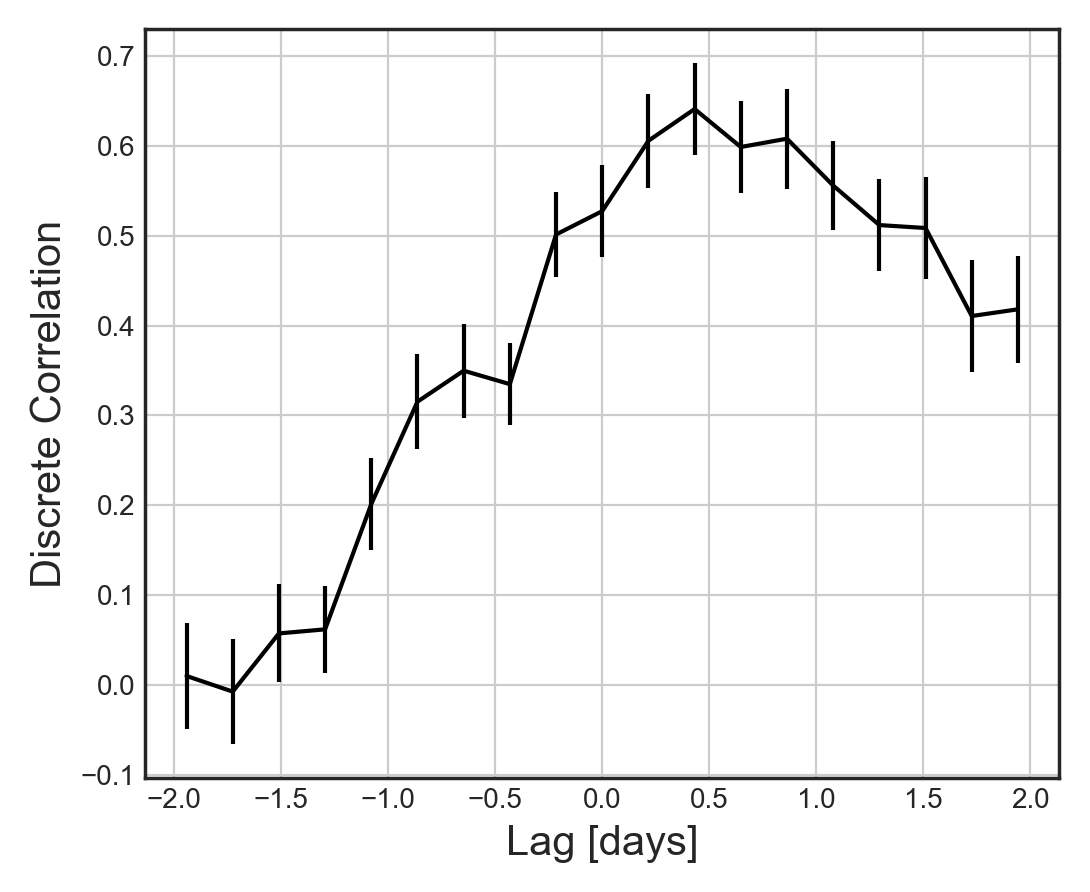}
    \includegraphics[width=0.33\linewidth]{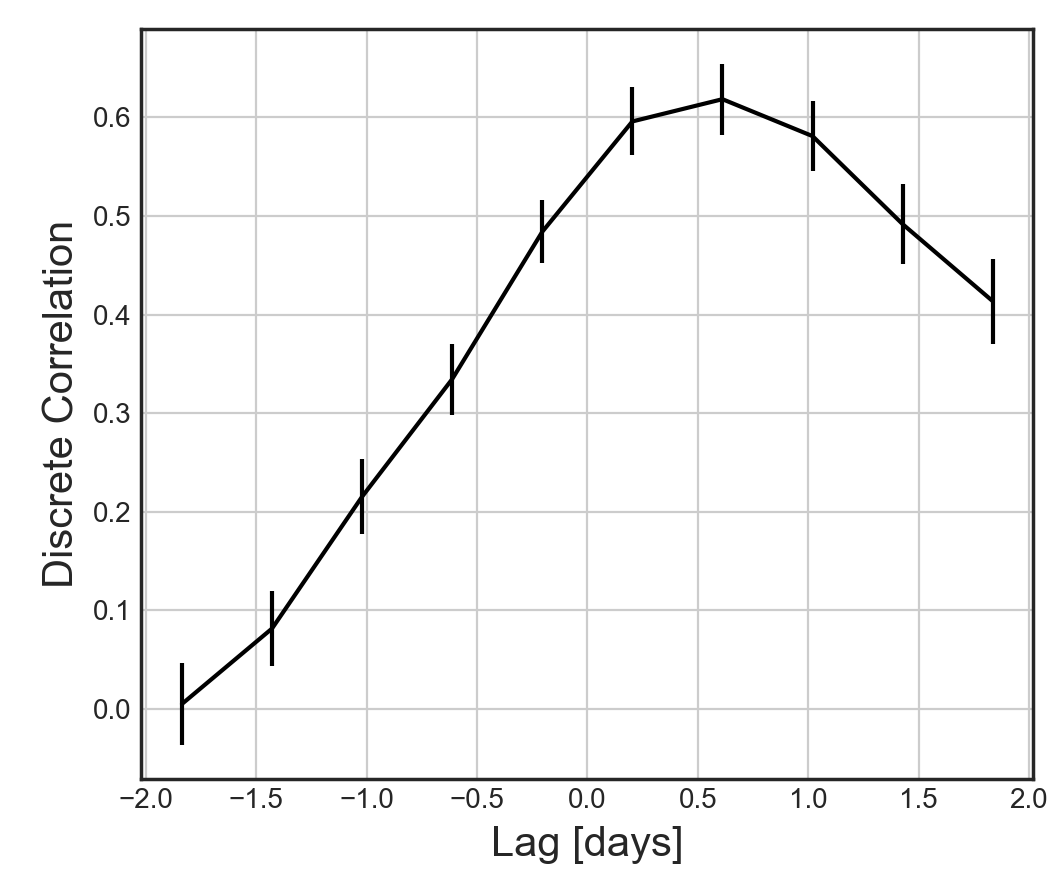}
}
\caption[General Short Caption]{The effect of the rolling window on the MICF estimate is shown here. The MICF with no window (Left:), one averaging over 2 consecutive points (Center:) and 4 consecutive points (Right:) all show peaks close to the predicted lag feature at 0.1 days.}
\label{fig:appendix-dcfxrayuvw2}
\end{figure*}
Just like the MICF, the DCF too is quite noisy on account of the stochasticity and the limitations of the observed lightcurves. As stated earlier, we use the DCF estimator of \cite{1988ApJ...333..646E} implemented in pyDCF \footnote{\protect\url{https://github.com/astronomerdamo/pydcf}}. This has an inbuilt option for different types of weighting and smoothing. We chose the uniform or {\it slot} weighting option over a window of 2 (center) and 4 ( right) points respectively and compare it with the unsmoothed DCF (left) in figure~\ref{fig:appendix-dcfxrayuvw2}. The comparison shows a decrease in the noisiness as the window size increases, with the largest window revealing the smoothest version of the principal lag detected in \cite{2018MNRAS.480.2881M} and confirmed by us. The peak position or the central value of the lag also shifts to greater values moving from no window to 4 point window. However this shift is not larger than the uncertainty on the lag itself. 

\nocite{*}
\bibliography{aapmsamp}
\end{document}